\documentclass[final,3p,times,twocolumn]{elsarticle}

\usepackage[pdftex]{color}
\usepackage{pgfplots}
\usepackage[font=footnotesize,labelfont=bf]{caption}
\usepackage[font=footnotesize,labelfont=bf]{subcaption}
\usepackage{multirow}
\usepackage{array}
\usepackage{longtable}
\usepackage{tikz}
\usetikzlibrary{positioning}

\usepackage{amssymb}

\journal{Signal Processing: Image Communication}

\begin{document}

\begin{frontmatter}



\title{MFCC-based Recurrent Neural Network for Automatic Clinical Depression Recognition and Assessment from Speech}


%


\author[1,2,3]{Emna Rejaibi}

\author[4]{Ali Komaty}

\author[5]{Fabrice Meriaudeau}

\author[3]{Said Agrebi}

\author[1]{Alice Othmani}
\ead{Corresponding author: alice.othmani@u-pec.fr}

\address[1]{Universit\'e Paris-Est, LISSI, UPEC, 94400 Vitry sur Seine, France}
\address[2]{ INSAT Institut National des Sciences Appliqu\'ees et de Technologie, Centre Urbain Nord BP 676-1080, Tunis, Tunisie}
\address[3]{Yobitrust, Technopark El Gazala B11
Route de Raoued Km 3.5,
2088 Ariana, Tunisie}
\address[4]{University of Sciences and Arts in Lebanon, Ghobeiry, Liban}
\address[5]{Universit\'e de Bourgogne Franche Comt\'e, ImvIA EA7535/IFTIM}

\begin{abstract}
Clinical depression or Major Depressive Disorder (MDD) is a common and serious medical illness. In this paper, a deep recurrent neural network-based framework is presented to detect depression and to predict its severity level from speech. Low-level and high-level audio features are extracted from audio recordings to predict the 24 scores of the Patient Health Questionnaire and the binary class of depression diagnosis. To overcome the problem of the small size of Speech Depression Recognition (SDR) datasets, expanding training labels and transferred features are considered. The proposed approach outperforms the state-of-art approaches on the DAIC-WOZ database with an overall accuracy of 76.27\% and a root mean square error of 0.4 in assessing depression, while a root mean square error of 0.168 is achieved in predicting the depression severity levels.
The proposed framework has several advantages (fastness, non-invasiveness, and non-intrusion), which makes it convenient for real-time applications. The performances of the proposed approach are evaluated under a multi-modal and a multi-features experiments. MFCC based high-level features hold relevant information related to depression. Yet, adding visual action units and different other acoustic features further boosts the classification results by 20\% and 10\% to reach an accuracy of 95.6\% and 86\%, respectively. Considering visual-facial modality needs to be carefully studied as it sparks patient privacy concerns while adding more acoustic features increases the computation time.

\end{abstract}




\end{frontmatter}

\section{Introduction}

Depression is a mental disorder caused by several factors: psychological, social or even physical factors. Psychological factors are related to permanent stress and the inability to successfully cope with difficult situations. Social factors concern relationship struggles with family or friends and physical factors cover head injuries. Depression describes a loss of interest in every exciting and joyful aspect of everyday life. Mood disorders and mood swings are temporary mental states taking an essential part of daily events, whereas, depression is more permanent and can lead to suicide at its extreme severity levels. Depression is a mood disorder that is persistent for up to eight months and beyond. According to the World Health Organization (WHO), 350 million people, globally, are diagnosed with depression. A recent study estimated the total economic burden of depression to be 210 billion US Dollars per year \cite{greenberg2015economic}, caused mainly by increased absenteeism and reduced productivity in the workplace. In many cases, the affected person denies facing mental disorders like depression, thus, he/she does not get the proper treatment.\\

Fortunately, depression is a curable disease. Physicians make clinical evaluations based on patients' self‐reports of their symptoms and standard mental health questionnaires such as the depression severity questionnaires. The depression severity assessment tests are multiple-choice self-report questionnaires that the patient takes. According to the answer, a score is automatically assigned. The Patient Health Questionnaire (PHQ) is a commonly used test and it is composed of nine clinical questions. The PHQ score assigned describes the depression severity level which ranges from 0 to 23. Even when patients self-report their symptoms, doctors correctly identify depression only half the time \cite{mitchell2009clinical}. This is mainly due to many similar symptoms between depression and other illnesses like hypothyroidism, hypoglycemia or even normal stress due to busy daily work. 

Recently, automatic mental states and mental disorders recognition have attracted considerable attention from computer vision and artificial intelligence community. Aiming to improve the human-machine interactions, several systems have been developed to automatically assess the emotions and the current mental state of a person \cite{dornaika2007inferring}. As human-machine communication takes place through audio-visual sensors, these developed systems study the best features to select from these two modalities to reach the best possible communication quality and boost performances \cite{wilkins2007inexpensive}. Several approaches have been developed so far to assess mental disorders such as depression more objectively based on external symptoms like facial expression, head movements, and speech. It has been shown that depression affects speech production \cite{scherer2015self}, more particularly it drops the range of pitch and volume, so the voice becomes softer and lower.

Numerous studies  in  the  literature  use  multi-modal fusion systems combining facial actions (visual cues), vocal prosody and text features \cite{williamson2016detecting}. Speech has been proven to be robust in the diagnosis of depression. In several works, depression prediction results using speech only outperform those using visual features or text \cite{williamson2016detecting}\cite{yang2017hybrid}. The fusion of the two modalities: audio and video leads to better depression prediction results \cite{williamson2016detecting}\cite{valstar2016avec}. But fusing three modalities: audio, video and text, decreases performances \cite{yang2017hybrid}. Speech can be measured cheaply, remotely, non-invasively and non-intrusively \cite{valstar2016avec} leading to a potent impact in recognizing depression. For this reason, the proposed study in this paper focuses on detecting depression using only speech recordings.

The use of various depression datasets and depression estimation approaches make it hard to decide which acoustic features show a better performance for the assessment of depression \cite{lopez2014study}. The Mel Frequency Cepstral Coefficients (the MFCCs) have proven their high efficiency in detecting depression compared to other audio features in shallow-based approaches \cite{lopez2014study}\cite{cummins2011investigation}. They are also considered as top audio features in speech-based applications like speech and speaker recognition \cite{tiwari2010mfcc}\cite{ma2016depaudionet}. \\

In this paper, low-level and high-level audio features are used in a deep neural network to assess depression. The Recurrent Neural Network is highly performing in speech recognition. This is why, the Short Long-Term Memory (LSTM) is chosen for extracting high-level audio features. The audio features to be trained within the network are the Mel Frequency Cepstral Coefficients and their first and second-degree derivatives. Throughout this study, depression is assessed by the self-report depression test of the Patient Health Questionnaire of eight questions (the PHQ-8). The aim of the study is to predict depression/non-depression by predicting the PHQ-8 binary score. A binary score of 0 is given to assess non-depression while a score of 1 is given when depression is diagnosed.
It also aims to assess the depression severity levels by predicting the 24 PHQ-8 scores (0 for non-depression; 10 for moderate depression; 23 for severe depression).\\

The outline of this paper is as follows. The related works are introduced in Section 2. Next, the proposed approach based on deep recurrent neural network is presented in Section 3. Section 4 illustrates and analyzes the experimental results and the depression corpus used. Finally, the conclusion and future works are provided in Section 5. 

\section{Related work}
Several works in automatic depression recognition and assessment are reported in the literature \cite{pampouchidou2017automatic}. Automatic depression recognition has become more and more popular since 2011 with the emergence of the eight successive editions of the Audio/Visual Emotion Challenge AVEC \cite{ringeval2018avec}.

Typically, depressed individuals tend to change their expressions at a very slow rate and pronounce flat sentences with stretched pauses \cite{pampouchidou2017detection}. Therefore, to detect depression, two types of features are frequently used: facial geometry features and audio features for their ability and consistency to reveal signs of
depression. The majority of approaches proposed, have the same structure with four main processing steps: preprocessing, feature extraction, dimension reduction, and classification \cite{jiang2017investigation}. \\

Two different approaches are mainly adopted to assess depression: hand-crafted features based approaches and deep-learning based approaches. Deep-learning based approaches outperform hand-crafted ones. The best reported performance for automatic depression recognition from speech to the best of our knowledge is presented in \cite{yang2017hybrid} where an approach based on 238 low-level audio features are fed to a Deep Convolutional Neural Network followed by a Deep Neural Network. The best root mean square error obtained reaches 1.46 over a group of depressed men in the prediction of the PHQ-8 scores.

\subsection{Hand-crafted features-based approaches} 

This family of approaches tackles two different tasks: hand-crafted audio features extraction and classification. The extracted hand-crafted audio features to detect depression might be classified into five main groups \cite{low2010detection} \cite{jiang2018detecting}: 

\begin{itemize}
  \item The \textbf{Spectral} features: related to the spectrum analysis like the \textit{Spectral Centroid} that locates the center of gravity of the spectrum, the \textit{Spectral Flatness} that determines the tone level of a band of the spectrum and the \textit{Energy}  \cite{low2010detection} \cite{jiang2018detecting} \cite{ringeval2015avec}. 
  
  \item The \textbf{Cepstral} features: related to the Cepstrum analysis (an anagram to the Spectrum signal) like the \textit{Mel Frequency Cepstral Coefficients (MFCCs)} that are considered to be the most commonly used audio features in speakers recognition for their high performance in describing the variation of low frequencies of the signal \cite{low2010detection} \cite{jiang2018detecting} \cite{alghowinem2013comparative} . 
  
  \item The \textbf{Glottis} features: derived from the vocal tract, the organ of the Human body responsible over producing speech \cite{low2010detection} \cite{jiang2018detecting}. 
  
  \item The \textbf{Prosodic} features: describe the speech intonation,  rate, and rhythm, like the \textit{Fundamental Frequency F0} (the first signal harmonic) and the \textit{Loudness} \cite{valstar2016avec} \cite{jiang2018detecting} \cite{ringeval2017avec}  . 
  
  \item The \textbf{Voice Quality}: like the \textit{Formants} (the spectrum maxima), the \textit{Jitter} (the signal fluctuation) and the \textit{Shimmer} (the peaks variation) \cite{valstar2016avec}\cite{ringeval2017avec}. 
\end{itemize}

The audio features extraction is processed for different segmentation window lengths. \cite{valstar2013avec} proposes better depression prediction results using the window of 20s, shifted forward by 1s, compared to a 3s-window. Meanwhile, \cite{yang2017hybrid} proposes a window of 60ms, shifted forward by 10ms, to predict depression using a neural network model.  \\

A comparative study of the performances of various classifiers in detecting depression from spontaneous speech was established in \cite{alghowinem2013comparative}. Three audio features fusion methods are tested for each classifier: features fusion, score fusion (one score is assigned for each feature while classifying it), and decision fusion (weighted majority voting). \\ 

Five classifiers are tested: the Gaussian Mixture Models (GMM), the Support Vector Machines (SVM) with raw data, the Support Vector Machines with GMM, the Multilayer Perceptron neural networks (MLP), and the Hierarchical Fuzzy Signature (HFS). The model of the SVM with GMM outperforms the other classifiers with the decision fusion method. The accuracy achieved is 81.61\%. The least performing classifier is the GMM with the features fusion method. The worst accuracy achieved with GMM is 48.26\%. 

\subsection{Deep learning-based approaches} 

The deep learning-based approaches for SDR could be categorized into two groups: those which use the raw audio signal as input and others which extract hand-crafted features and use them as input of the deep neural network. For instance, \cite{ma2016depaudionet} and \cite{trigeorgis2016adieu} propose a deep learning-based approaches that use the raw audio signal as input. While, the Mel-Scale Filter Bank is applied in \cite{ma2016depaudionet} and the extracted features are fed to a Convolutional Neural Network (CNN). \\

Different studies compare different deep learning-based architectures such as the Deep Convolutional Neural Network (DCNN), the Deep Convolutional Neural Network followed by a Deep Network (DCNN-DNN) and the Long Short Term Memory network (LSTM). The DCNN-DNN outperforms the DCNN in predicting the PHQ-8 scores of depression severity levels \cite{yang2017multimodal}. The best results achieved with DCNN-DNN is a root mean square error of 1.46 on a group of depressed men \cite{yang2017hybrid}. The performance of the LSTM in predicting depression is evaluated with the F1 score. The recurrent neural network proposed achieves only 52\% in detecting depression but reaches 70\% when it comes to detecting non-depression \cite{ma2016depaudionet}. \\

The deep learning based approaches highlight the fact that the gender (male/female) has an impact on the model's performances. The depression assessment results in \cite{yang2017hybrid} show that the root mean square error achieved with the DCNN-DNN model on a group of depressed women is three times higher that the root mean square error achieved on a group of depressed men using the same model. 

\section{Proposed Method}

\subsection{Method overview}
The deep learning based approach proposed in this paper aims to assess depression and predict its severity levels using the Mel Frequency Cepstral Coefficients and the Long Short Term Memory network. \\

\begin{figure*}
   \centering 
   \includegraphics[width=.9\textwidth]{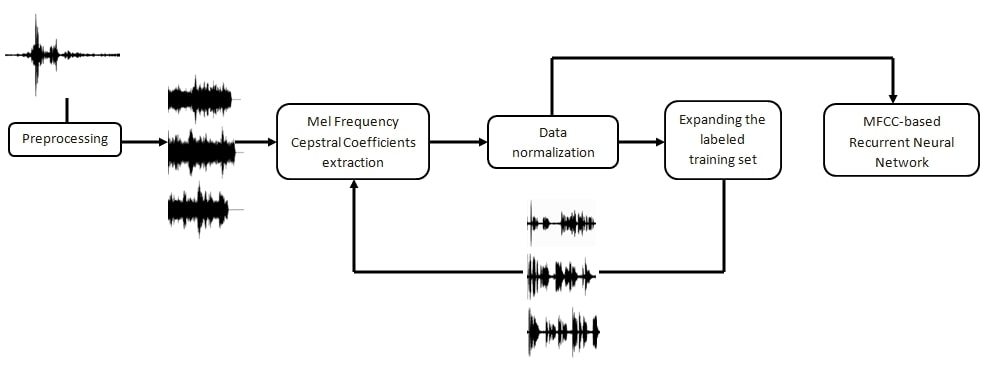}
   \caption{The proposed approach to assess depression. Fist, the audio recordings of the clinical interviews are preprocessed and the audio segments of the participants' speech only are retrieved. The low-level features are then extracted from the audio segments and normalized. The labeled training set is expanded through transfer learning and data augmentation where new audio segments of the participants' speech only are generated. Finally, the MFCC-based Recurrent Neural Network is trained to detect depression/non-depression or to predict the depression severity level.}   
   \label{approach} 
\end{figure*}

The steps followed throughout this study are summarized in Fig.~\ref{approach}. First, the audio signals are preprocessed (Section~\ref{preprocessing}). Next, the low-level audio descriptors are extracted and normalized (Section~\ref{mfcc_extraction} and Section~\ref{mfcc_normalization}). The low-level features are the MFCC features. In the following step, these MFCC features are fed to the deep neural network for depression prediction (see Section~\ref{mfcc_based_rnn}). Depression datasets available to assess depression from speech are relatively small. To overcome this challenge, data augmentation is performed and described in Section~\ref{data_augmentation} and knowledge transfer from related task is eventually carried out in Section~\ref{transfer_learning}.
The proposed deep-based framework is presented with more details in the following contents. 

\subsubsection{Preprocessing} 
\label{preprocessing}
The audio recordings are clinical interviews. They are conversations between an interviewer and the participants (the interviewees). The recordings are preprocessed in order to retrieve the speech of the participants only. As the main goal is to automatically detect depression from the participants' spoken answers, the recordings are separated into two groups by the speaker: one group has the audio segments of the participants and the other has the audio segments of the interviewer. The audio segments of the interviewer are no longer used in this study.

\subsubsection{Low-Level Features Extraction}
\label{mfcc_extraction}
The Mel Frequency Cepstral Coefficients (MFCC) are the most commonly used audio features in speaker recognition due to their robustness in describing the variation of low frequencies signal. The MFCC coefficients describe the energies of the cepstrum in a non-linear scale, the mel-scale. They are considered as the most discriminative acoustic features that approximate how the "human peripheral auditory system" perceives the speech signal \cite{tiwari2010mfcc}. The first and the second derivatives of these coefficients allow to track their variation over time and thus track the variation of the speech tone \cite{janse2014comparative}. For these reasons, in this proposed work, only the MFCC coefficients are extracted in order to study their robustness in a speech-based application of automatic diagnosis of depression. 

In this work, the low-level features are defined as the MFCC coefficients and they are extracted from the preprocessed audio recordings.
The speech signal is first divided into frames by applying a windowing function of 2.5s at fixed intervals of 500 ms. The Hamming window is used as window function to remove edge effects.\\

A cepstral feature vector is then generated for each frame. The Discrete Fourier Transform (DFT) is computed for each frame. Only the logarithm of the amplitude spectrum is retained. The spectrum is after smoothed to emphasize perceptually meaningful frequencies. 24 spectral components into 44100 frequency bins are collected in the Mel frequency scale. The components of the Mel-spectral vectors calculated for each frame are highly correlated. Therefore, the Karhunen-Loeve (KL) transform is applied to the Mel-spectral vectors to decorrelate their components. The KL transform is approximated by the Discrete Cosine Transform (DCT). Finally, 60 cepstral features are obtained for each frame. 

\subsubsection{Data Normalization} 
\label{mfcc_normalization}
Since the range values of the MFCC coefficients vary widely, their impact within the deep network might be non-uniform and the gradient descent might converges to null very fast. Therefore, the range values of all the MFCC coefficients should be normalized beforehand. \\

The Sandardization method (the Z-score Normalization) is the most commonly used scaling method with audio features. As the same channel conditions are used for all the audio recordings, the MFCC coefficients are not normalized per 60ms-audio frames. They are, rather, normalized all at once using the mean value and the standard deviation. The distribution of the mean and the standard deviation are calculated across the coefficients. Next, the mean is substracted from each Mel scale frequency coefficient, that is, later, divided by the standard deviation.

\subsubsection{High-Level Features Extraction and Classification}
\label{mfcc_based_rnn}
\begin{figure*}[]
   \centering
   \includegraphics[width=0.92\textwidth]{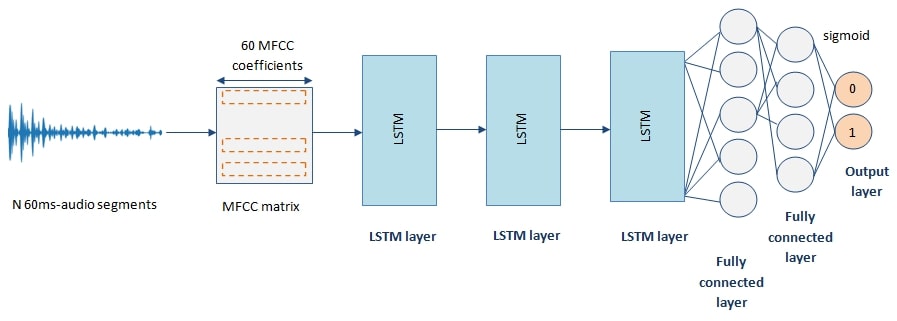}
   \caption{Overall structure of the MFCC-based RNN approach for automatic depression recognition. After preprocessing an audio sequence, an MFCC matrix of size n audio frames of 60ms by the 60 Mel Frequency Cepstral Coefficients is generated and then fed to the MFCC-based RNN. The MFCC-based RNN presents three LSTM layers followed by two fully connected layers. To predict the PHQ-8 binary  (0 for non-depression and 1 for depression), the output layer is a dense layer of size 2 neurones with a sigmoid activation function. While, to predict the PHQ-8 scores (the depression severity level), the output layer is a dense layer of size 24 neurones with a softmax activation function. }
   \label{figure_8} 
\end{figure*}

An MFCC-based RNN is proposed as a high-level features extractor and classifier as shown in Fig.~\ref{figure_8}. As LSTM is one of the most performing recurrent neural networks, it is used in the baseline of the proposed model. The input of the deep model is the MFCC extracted matrix of size n (n is the total number of 60ms audio frames extracted from the audio segments) by the 60 Mel Frequency Cepstral Coefficients.  
The architecture proposed is composed of three successive LSTM layers followed by two Dense layers as shown in Fig.~\ref{figure_8}.\\

The model predicts depression and assesses its severity level. The output layer of the MFCC-based RNN is a two-cell dense layer activated with a sigmoid function to predict the PHQ-8 binary. However, the output layer to predict the PHQ-8 scores is a dense layer of 24 neurons activated with a softmax function.

\subsubsection{Data augmentation}
\label{data_augmentation}
Aiming to improve the MFCC-based RNN's performances and to avoid overfitting, data augmentation is carried out to increase and to diversify the input data. Data augmentation techniques are applied on the preprocessed audio segments of each participant’s speech.\\

New audio segments are generated by applying the following four different data augmentation techniques over the preprocessed audio segments:

\begin{itemize}
  \item \textbf{Noise Injection} : adds random values into the data with a noise factor of 0.05.
  \item \textbf{Pitch Augmenter} : randomly changes the pitch of the signal. The pitch factor is 1.5. 
  \item \textbf{Shift Augmenter} : randomly shifts the audio signal to left or right. The shift max value used is 0.2 seconds. If the audio signal is shifted to the left with x seconds, the first x seconds are marked as silence (Fast Forwarding the audio signal). If the audio signal is shifted to the right with x seconds, the last x seconds are marked as silence (Back Forwarding the audio signal). 
  \item \textbf{Speed Augmenter} : stretches times series by a fixed rate with a speed factor of 1.5. 
\end{itemize}

\begin{figure}
   \begin{subfigure}[]{\columnwidth}
         \includegraphics[width=\textwidth]{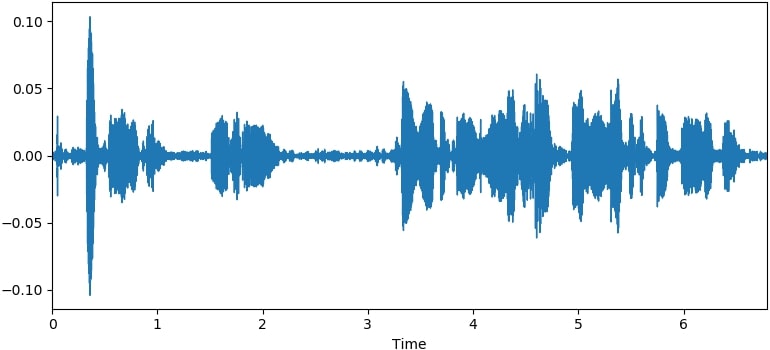}
         \caption{The preprocessed audio segment before data augmentation.}
         \label{signal_1}
   \end{subfigure}
   \begin{subfigure}[]{\columnwidth}
         \includegraphics[width=\textwidth]{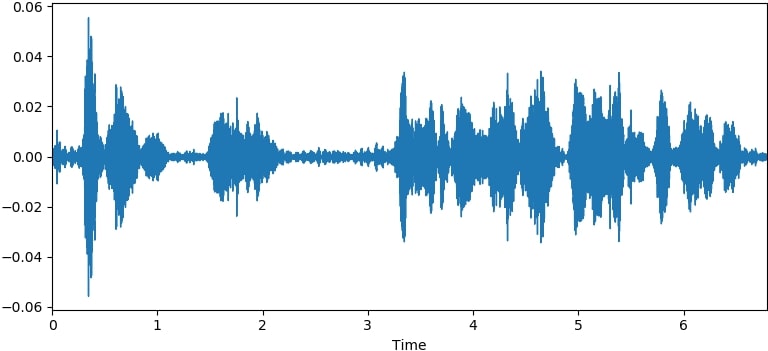}
         \caption{The generated audio segment after pitch augmentation.}
         \label{signal_2}
   \end{subfigure}  
   \caption{An example of a preprocessed audio segment of one participant’s speech before and after data augmentation. The data augmentation technique used is the random pitch augmenter with a pitch factor of 1.5. }
   \label{signal}     
\end{figure}

Fig.~\ref{signal} displays an example of a random participant' preprocessed audio segment before and after data augmentation. The data augmentation technique performed in this example is the random pitch augmenter with a pitch factor of 1.5.

\subsubsection{Transfer Learning}
\label{transfer_learning}
The core challenge to overcome throughout this study is the limited depression data available as an input to the MFCC-based RNN. The second proposed solution is to transfer knowledge from an independent, yet related, previously learned task.\\ 

For that, a pretraining is first applied to pretrain the MFCC-based RNN on a related task. In this study, emotions recognition from speech is chosen as the related task. Second, fine-tuning on the target task, depression recognition from speech, is performed. The number of neurones of the third dense layers in the MFCC-based RNN model is modified and it is chosen to be as the number of emotions in the related task dataset. Once the MFCC-based RNN is pretrained on the emotions recognition dataset, the optimum weights of the model are used as a starting point during the fine-tuning of the model on depression recognition dataset. 

\section{Experiments and results}
\subsection{Datasets}

Three datasets have been used in the experiments:

\subsubsection{DAIC-WOZ corpus}

The main dataset used in this paper to assess depression is the DAIC-WOZ depression dataset \cite{gratch2014distress} which is used in the AVEC2017 challenge \cite{ringeval2017avec}.  \\
The DAIC-WOZ corpus is designed to support the diagnosis of psychological distress conditions: depression, post traumatic stress disorder (PTSD), etc. It provides audio recordings of 189 clinical interviews of 189 participants answering the questions of an animated virtual interviewer named Ellie. Each recording is labeled by the PHQ-8 score and the PHQ-8 binary. The PHQ-8 score defines the severity level of depression of the participant and the PHQ-8 binary defines whether the participant is depressed or not. For technical reasons, only 182 audio recordings are used. The average length of the recordings is 15 minutes with a fixed sampling rate of 16 kHz.\\

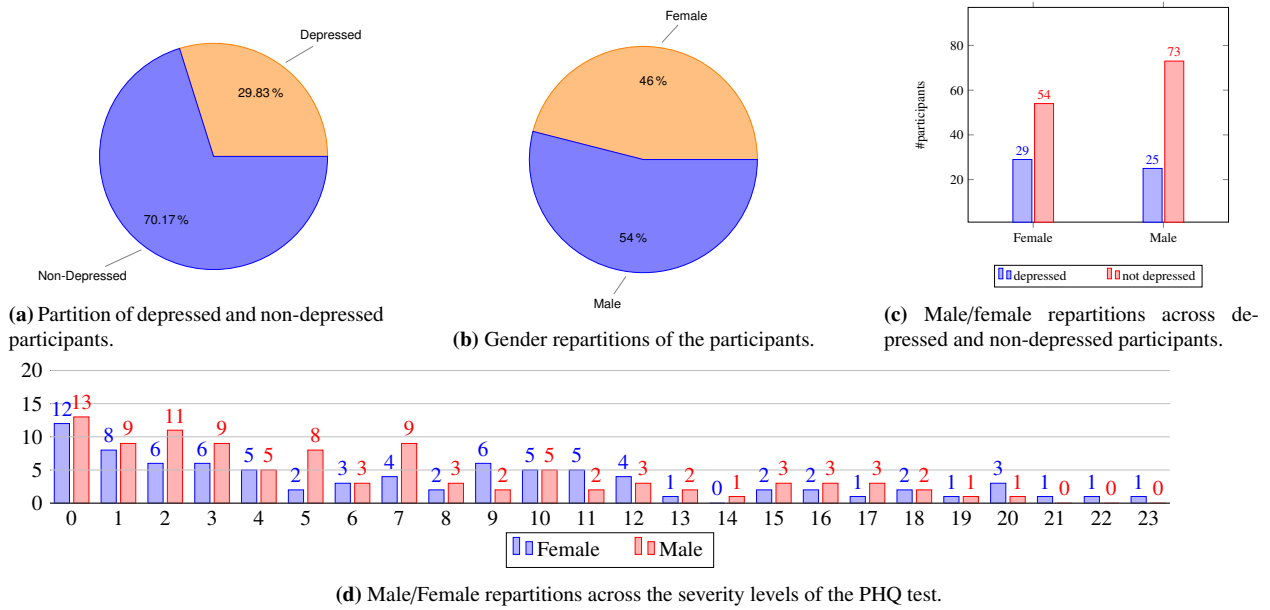
\begin{figure*}[]
   \centering 
   \begin{subfigure}[b]{0.3\textwidth}
         \centering 
         \scalebox{0.5}{\def\angle{0}
\def\radius{3}
\def\cyclelist{{"orange","blue","red","green"}}
\newcount\cyclecount \cyclecount=-1
\newcount\ind \ind=-1
\begin{tikzpicture}[nodes = {font=\sffamily}]
  \foreach \percent/\name in {
      29.83/Depressed,
      70.17/Non-Depressed 
    } {
      \ifx\percent\empty\else               
        \global\advance\cyclecount by 1     
        \global\advance\ind by 1            
        \ifnum3<\cyclecount                 
          \global\cyclecount=0              
          \global\ind=0                     
        \fi
        \pgfmathparse{\cyclelist[\the\ind]} 
        \edef\color{\pgfmathresult}         
        \draw[fill={\color!50},draw={\color}] (0,0) -- (\angle:\radius)
          arc (\angle:\angle+\percent*3.6:\radius) -- cycle;
        \node at (\angle+0.5*\percent*3.6:0.7*\radius) {\percent\,\%};
        \node[pin=\angle+0.5*\percent*3.6:\name]
          at (\angle+0.5*\percent*3.6:\radius) {};
        \pgfmathparse{\angle+\percent*3.6}  
        \xdef\angle{\pgfmathresult}         
      \fi
    };
\end{tikzpicture}}
         \caption{Partition of depressed and non-depressed participants.}
         \label{figure_3}
   \end{subfigure}
    \hspace*{\fill}
    \begin{subfigure}[b]{0.3\textwidth}
       \centering 
       \scalebox{0.5}{\def\angle{0}
\def\radius{3}
\def\cyclelist{{"orange","blue","red","green"}}
\newcount\cyclecount \cyclecount=-1
\newcount\ind \ind=-1
\begin{tikzpicture}[nodes = {font=\sffamily}]
  \foreach \percent/\name in {
      46/Female,
      54/Male
    } {
      \ifx\percent\empty\else               
        \global\advance\cyclecount by 1     
        \global\advance\ind by 1            
        \ifnum3<\cyclecount                 
          \global\cyclecount=0              
          \global\ind=0                     
        \fi
        \pgfmathparse{\cyclelist[\the\ind]} 
        \edef\color{\pgfmathresult}         
        \draw[fill={\color!50},draw={\color}] (0,0) -- (\angle:\radius)
          arc (\angle:\angle+\percent*3.6:\radius) -- cycle;
        \node at (\angle+0.5*\percent*3.6:0.7*\radius) {\percent\,\%};
        \node[pin=\angle+0.5*\percent*3.6:\name]
          at (\angle+0.5*\percent*3.6:\radius) {};
        \pgfmathparse{\angle+\percent*3.6}  
        \xdef\angle{\pgfmathresult}         
      \fi
    };
\end{tikzpicture}}
       \caption{Gender repartitions of the participants.}
       \label{figure_5}
    \end{subfigure}
    \hspace*{\fill}
     \begin{subfigure}[b]{0.3\textwidth}
       \centering 
       \scalebox{0.5}{\begin{tikzpicture}
\begin{axis}[
    ybar,
    enlargelimits=0.5,
    legend style={at={(0.5,-0.15)},
      anchor=north,legend columns=-1},
    ylabel={\#participants},
    symbolic x coords={Female,Male},
    xtick=data,
    bar width=0.5cm,
    nodes near coords,
    nodes near coords align={vertical},
    legend style={
            at={(0.5,-0.2)},
            anchor=north,
            legend columns=-1,
            /tikz/every even column/.append style={column sep=1cm}
        },
    ]
\addplot coordinates {(Female,29) (Male,25)};
\addplot coordinates {(Female,54) (Male,73)};
\legend{depressed,not depressed}
\end{axis}
\end{tikzpicture}}
       \caption{Male/female repartitions across depressed and non-depressed participants.}
       \label{figure_6} 
     \end{subfigure}
    \hspace*{\fill} 
    \begin{subfigure}[b]{1\textwidth}
       \scalebox{0.8}{\begin{tikzpicture}
  \centering
  \begin{axis}[
        ybar, axis on top,
        height=4cm, width=20cm,
        bar width=0.25cm,
        ymajorgrids, tick align=inside,
        enlarge y limits={value=.1,upper},
        enlarge x limits=0.02,
        ymin=0, ymax=20,
        axis x line*=bottom,
        axis y line*=left,
        y axis line style={opacity=0},
        tickwidth=0pt,
        legend style={
            at={(0.5,-0.2)},
            anchor=north,
            legend columns=-1,
            /tikz/every even column/.append style={column sep=0.5cm}
        },
        symbolic x coords={0,1,2,3,4,5,6,7,8,9,10,11,12,13,14,15,16,17,18,19,20,21,22,23},
       xtick=data,
       nodes near coords={
        \pgfmathprintnumber[precision=0]{\pgfplotspointmeta}
       }
    ]
    \addplot coordinates {(0,12) (1,8) (2,6) (3,6) (4,5) (5,2) (6,3) (7,4) (8,2) (9,6) (10,5) (11,5) (12,4) (13,1) (14,0) (15,2) (16,2) (17,1) (18,2) (19,1) (20,3) (21,1) (22,1) (23,1)};
\addplot coordinates {(0,13) (1,9) (2,11) (3,9) (4,5) (5,8) (6,3) (7,9) (8,3) (9,2) (10,5) (11,2) (12,3) (13,2) (14,1) (15,3) (16,3) (17,3) (18,2) (19,1) (20,1) (21,0) (22,0) (23,0)};
\legend{Female,Male}
  \end{axis}
  \end{tikzpicture}}
       \caption{Male/Female repartitions across the severity levels of the PHQ test.}
       \label{figure_7}
    \end{subfigure}
\caption{Gender, depression and severity level repartitions of the participants within the DAIC-WOZ Corpus. (a) Depressed versus Non-Depressed participants. The PHQ-8 binary of Depressed participants is 1 and the PHQ-8 binary of Non-Depressed participants is 0. (b) Gender repartition of the participants. (c) The repartition of males and females across the depressed and the non-depressed participants. Blue color for depressed while pink color for non-depressed. (d) Male and female participants repartitions across the 24 depression severity levels given by the PHQ-8 test. }    
\label{repartition}
\end{figure*}

The repartitions of the participants by their gender, depression, and depression severity level are shown in Fig.~\ref{repartition}. Almost half of the participants are females (46\%) (Fig.~\ref{figure_5}) and the third of the participants are labeled depressed (Fig.~\ref{figure_3}). Among the depressed participants, almost half of them are females (29 out of 54 participants) (Fig.~\ref{figure_6}). According to the repartitions in Fig.~\ref{repartition}, the dataset is gender-balanced. However, it is class-imbalanced as the number of non-depressed participants is three times higher than the number of depressed participants. After data preprocessing, 80\% of the audio segments are used for training, 10\% of them are used for validation and 10\% for testing.  

\subsubsection{RAVDESS dataset}

The experiment of transfer learning described in Section~\ref{transfer_learning}, uses a related task dataset. In this study, RAVDESS dataset \cite{livingstone2018ryerson} is used for this purpose. It is a dataset for emotions recognition \cite{ouyang2017audio, atalay2018comparison, pham2018end}. It contains audio recordings of 24 actors expressing eight different emotions: neutral, calm, happy, sad, angry, fearful, disgust, and surprised. For each actor, eight trials per emotion are recorded for two different tasks: speaking and singing. The average length of the audio recordings within the dataset is five seconds. 
\subsubsection{AVi-D dataset}

The generalization of the proposed model to other dataset is eventually evaluated in Section~\ref{generalization_evaluation}. For that, the performance of trained MFCC-based RNN is evaluated on the AVi-D corpus which was introduced during the AVEC2014 challenge \cite{valstar2013avec, valstar2014avec}. 
The AVi-D dataset is depression and affects database. 300 audio recordings are collected from 292 participants who go through two different tasks: the Northwind task (reading task) and the Freeform task (answering questions). Approximately, 150 audio recordings are available per task. \\
The AVi-D corpus is based on the commonly used depression assessment test: The Beck Depression Inventory-II for assessing the severity levels of depression. The audio recordings are labeled by the BDI-II scores ranging from 0 to 63. A BDI-II score of 14 is the threshold to use for depression assessment. For a score below this threshold, the patient is labeled non-depressed. 

\subsection{Implementation details}
\paragraph{MFCC-based RNN implementation}
The MFCC matrix is of size n (n is the total number of 60ms audio frames extracted from the preprocessed audio segments) by the 60 Mel Frequency Cepstral Coefficients. Successive 60-unit input vectors are fed to the first LSTM layer. The three LSTM layers have 40, 30 and 20-output cell units. Each LSTM layer is parametrized as follows: 
\begin{itemize}
  \item the LSTM is activated with the hyperbolic tangent activation function (tanh), 
  \item the LSTM recurrent step is activated with the hard sigmoid activation function, 
  \item a recurrent dropout of 0.2\% is applied to prevent the recurrent state from overfitting, 
  \item the kernel weights are initialized using the glorot uniform initializer, 
  \item during the optimization, penalties are applied over the bias vector using the regularizer function to improve performances. The l1 imposed constraint is 0.001. 
\end{itemize}

A batch normalization layer is assigned to each LSTM layer along with 0.2\% dropout. The following two dense layers are of size 15 and 10, respectively. They are activated with the hyperbolic tangent function. 

The output dense layer is of size 2 and is activated with a sigmoid function to predict the PHQ-8 binary (0 for non-depression and 1 for depression). The sixth layer of the MFCC-based RNN model is replaced by an output dense layer of size 24 neurones with a softmax activation function to predict the 24 depression severity levels (the PHQ-8 scores).  

The optimizer used within the proposed deep framework is the Adam optimizer with a learning rate of $10^{-3}$ and a decay of $10^{-6}$. To prevent the model from training instability or training failure caused by a large learning rate or a tiny one, the learning rate used is an adaptive one. It is updated each epoch and decreases from $10^{-3}$ up to $10^{-10}$ according to the estimated error. To compare the proposed approach performances with previous works, the loss function calculated across the epochs is the root mean square error instead of the cross-entropy loss function.

The batch size for the training model across all the experiments is set to 130. After evaluating the results obtained with several batch sizes ranging from 32 to 500, the best size range  to be used is between 100 and 170; under 100 the model overfits  and beyond 170 the model underfits. 

\paragraph{MFCC-based RNN implementation in transfer learning}

The proposed framework is pretrained with the emotions recognition task. The last layer of the dense is of size eight neurones to predict the eight emotions classes. After pretraining, the model is fine-tuned with the target task of depression assessment: the first three LSTM layers are frozen and the last three Dense layers are retrained. 

\paragraph{Computational Complexity}
To run the experiments described above, the machine used is connected to an NVIDIA GeForce GTX 1080 GPU with 16GB RAM and 350GB storage. The operating system is windows 10 with its latest updates installed. Training the MFCC-based RNN takes five days. The prediction time is $2\ast10^{-3}$ seconds. The computational time of the proposed model is reasonable and satisfies real-time applications.

\subsection{Experimental results}

\subsubsection{MFCC-based RNN Evaluation}
\begin{figure}[]
   \begin{subfigure}[]{\columnwidth}
         \centering
         \includegraphics[width=\textwidth]{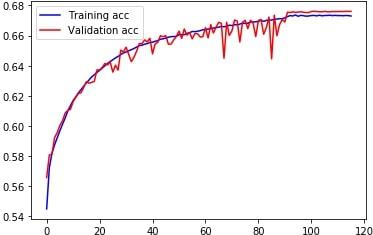}
         \caption{Training and Validation Accuracy.}
         \label{main_net_performance_1}
   \end{subfigure}
   \begin{subfigure}[]{\columnwidth}
         \centering
         \includegraphics[width=\textwidth]{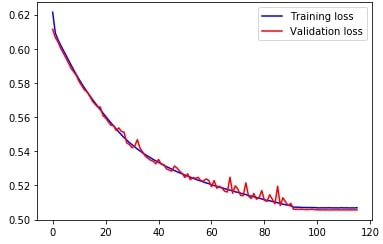}
         \caption{Training and Validation Loss.}
         \label{main_net_performance_2}
   \end{subfigure}   
   \caption{MFCC-based RNN accuracy and loss in the training and in the validation. A batch size of 130 over 120 epochs (displayed on the X-axis) is considered. DAIC-WOZ dataset is randomly divided into 80\% for training and 20\% for validation. During the training phase, 90\% of the training set is used for learning the weights and 10\% is used for testing. }
   \label{main_net_performance}     
\end{figure}

The proposed deep framework is trained to deliver two different tasks. The first task is to assess depression under the PHQ-8 binary test. The second one is to predict its severity levels under the PHQ-8 scores test. The obtained results are reported per sample (one sample corresponds to the MFCC coefficients extracted from 60ms-audio frame). The performance of the network is evaluated for both tasks as follows: 

\paragraph{Evaluation of the depression assessment task}
Table.~\ref{table_binary} summarizes the resulting performances after training the main network over the training and validation sets. The best validation accuracy achieved by the proposed architecture is 67.61\% with a root mean square error of 0.5 calculated on the validation set.
The model's performances over the training epochs are shown in Fig.~\ref{main_net_performance}. At the end of the training, the validation accuracy slightly exceeds the training accuracy (Fig.~\ref{main_net_performance_1}) which demonstrates that the MFCC-based RNN does not overfit at the end of the training and has the ability to make correct predictions on a new set of data unused for training. 

\paragraph{Evaluation of the depression severity levels prediction task}
The model is trained to predict 24 classes instead of two and the root mean square error (RMSE) achieved is three times better than the RMSE achieved in the depression assessment task. The RMSE reaches 0.168. 
The proposed approach performs ten times better than the benchmark used in this study in the depression severity levels prediction task. 

\subsubsection{Data augmentation evaluation} 
\label{data_augmentation_evaluation}
The data augmentation experiment is only evaluated over the depression assessment task. After training the MFCC-based RNN on the augmented dataset, the validation accuracy increased by 6.39\% to reach 74\%. While, the validation loss dropped by 0.08 and reached an RMSE of 0.42 (Table.~\ref{table_data_augmentation}). The performances improvement is explained by the more important number of input data and their diversification. \\ 

The confusion matrix in Fig.~\ref{confusion_matrices_DA} shows that 66\% of the samples are correctly identified as non-depression. But, only, 7.9\% were correctly identified as depression. 
Taking into consideration that the input data is imbalanced and that only the third of the participants are labeled as depressed (Fig.~\ref{figure_3}), the F1 score should be evaluated. The F1 score is based on the harmonic mean that punishes the extreme values caused by imbalanced data.
Table.~\ref{table_data_augmentation_matrix} shows that 70\% of the samples predicted as depression are well-classified. However, due to limited data, the ability to detect depression is 26\%. The F1 score for depression increases up to 38\% which proves that the network would improve more if more input data of depressed participants is introduced. As two thirds of the input data is for non-depressed participants the F1 score for non-depression reaches 84\%. 

\begin{figure*}[]
\centering
   \begin{subfigure}[]{.75\textwidth}
   \centering
         \includegraphics[width=\linewidth]{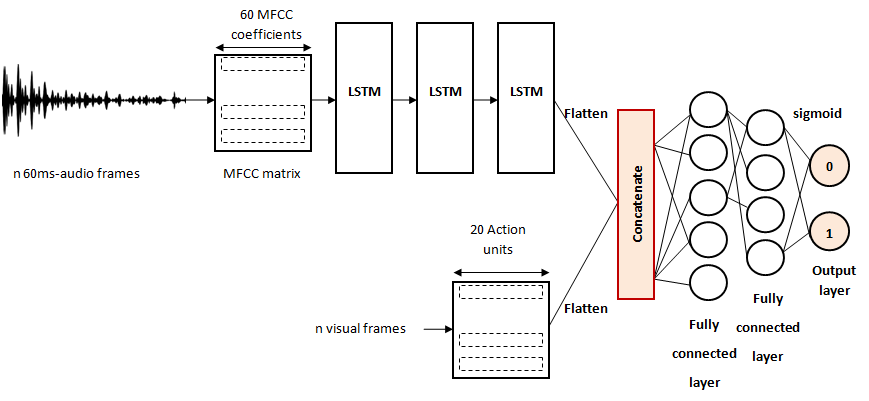}
         \caption{The Multi-Modal Experiment}
         \label{multi_modal}
   \end{subfigure}
   \begin{subfigure}[]{.75\textwidth}
    \centering
         \includegraphics[width=\linewidth]{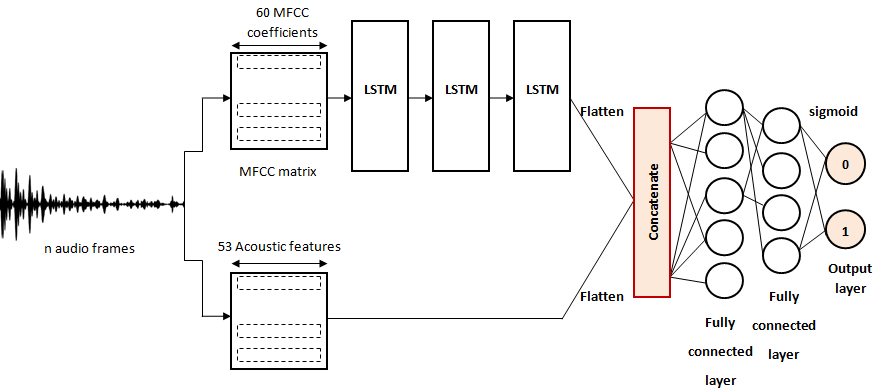}
         \caption{The Multi-Channel Experiment}
         \label{multi_channel}
   \end{subfigure}   
   \caption{Overall structure of the MFCC-based RNN approach for the depression assessment task over two different experiments: a)The Multi-Modal Experiment: 20 facial action units provided by the DAIC-WOZ corpus are concatenated with the flattened vector of the third LSTM layer's output. The new features' vector is then fed to the fully connected layers for classification.
   b)The Multi-Channel Experiment: 53 acoustic features are extracted from audio frames windowed by 10ms, they are concatenated with the MFCC-based high-level features and then fed to the fully connected layers.}
   \label{multi_modal_channel}     
\end{figure*}

\begin{table}[]
    \centering
    \begin{center}
        \begin{tabular}{|c|c|c|}
            \hline
            Results & Accuracy & RMSE  \\
            \hline
            Training Set & 67.33\%  & 0.5068 \\ 
            \hline
            Validation Set & 67.61\%  & 0.5057 \\
            \hline
        \end{tabular}
    \end{center}
    \caption{MFCC-based RNN performance in the PHQ-8 binary classification for the depression assessment task.}
    \label{table_binary}
\end{table}
\begin{table}[]
    \centering
    \begin{center}
        \begin{tabular}{|c|c|c|}
            \hline
            Validation Results & Accuracy & RMSE  \\
            \hline
            Female & 85\%  & 0.32\\ 
            \hline
            Male & 83\% & 0.34\\ 
            \hline
        \end{tabular}
    \end{center}
    \caption{Gender Effect over the MFCC-based RNN performance in the PHQ-8 binary classification.}
    \label{table_gender_effect}
\end{table}
\begin{table}[]
    \centering
    \begin{center}
        \begin{tabular}{|c|c|c|}
            \hline
                Results & Accuracy & RMSE  \\
            \hline
                Training Set & 74.66\%  & 0.4163 \\ 
            \hline
                Validation Set & 74\%  & 0.4206  \\
            \hline
        \end{tabular}
    \end{center}
    \caption{MFCC-based RNN performance in the PHQ-8 binary classification after data augmentation.}
    \label{table_data_augmentation}
\end{table}
\begin{table}[]
    \centering
    \begin{center}
        \begin{tabular}{|c|c|c|c|}
            \hline
                Class & Precision & Recall &  F1 Score \\
            \hline
                Non-Depressed & 75\%  & 95\% & 84\% \\ 
            \hline
                Depressed & 70\%  & 26\% & 38\% \\ 
            \hline
                Accuracy &  &  & 74\% \\  
            \hline
        \end{tabular}
    \end{center}
    \caption{MFCC-based RNN performances under depressed and non-depressed groups based on three metrics (precision, recall and FI score) after data Augmentation.}
\label{table_data_augmentation_matrix}
\end{table}

\begin{table}[]
    \centering
    \begin{center}
        \begin{tabular}{|c|c|c|}
            \hline
                Results & Accuracy & RMSE  \\
            \hline
                Training Set & 77.21\%  & 0.3991 \\ 
            \hline
                Validation Set & 76.27\%  & 0.4055 \\ 
            \hline
        \end{tabular}
    \end{center}
    \caption{MFCC-based RNN performances in the PHQ-8 binary classification after transfer learning setup. The proposed network is pretrained on the RAVDESS dataset and fine-tuned on the DAIC-WOZ dataset. }
\label{table_transfer_binary}
\end{table}

\begin{table}[]
    \centering
    \begin{center}
        \begin{tabular}{|c|c|c|c|}
            \hline
                Class & Precision & Recall &  F1 Score \\
            \hline
                Non-Depressed & 78\%  & 94\% & 85\% \\ 
            \hline
                Depressed & 69\%  & 35\% & 46\% \\ 
            \hline
                Accuracy &   &  & 76\% \\ 
            \hline
        \end{tabular}
    \end{center}
    \caption{MFCC-based RNN performances under depressed and non-depressed groups based on three metrics (precision, recall and FI score) after transferring knowledge from an other task.}
\label{table_transfer_confusion_matrix}
\end{table}

\begin{figure*}[]
	\begin{subfigure}[b]{0.35\textwidth} 
		\scalebox{0.6}{\definecolor{forestgreen}{rgb}{0.0, 0.27, 0.13}

 \begin{tikzpicture}[align=center,
box/.style={draw,rectangle,minimum size=2cm,text width=1.5cm,align=left},
box_green/.style={draw,rectangle,fill=green!50,minimum size=2cm,text width=1.5cm,align=left},
box_red/.style={draw,rectangle,fill=red!40,minimum size=2cm,text width=1.5cm,align=left}]
\matrix (conmat) [row sep=.1cm,column sep=.1cm] {
    \node (box11) [box_green,
    label=left:\textbf{Non-Depressed },
    label=above:\textbf{\textbf{Non-Depressed }},
    ] {131766 \\ 66.14\%};
    &
    \node (box12) [box_red,
    label=above:\textbf{Depressed },
    label=above right:\textbf{Sum\_lin},
    label=right: 176596\\ \textcolor{forestgreen}{74.61\%} \\ \textcolor{red}{25.39\%},
    ] {44830 \\ 22.5\%};
    \\
    \node (box21) [box_red,
    label={left:\textbf{Depressed }},
    label=below left:\textbf{Sum\_col},
    label=below:138645\\ \textcolor{forestgreen}{95.05\%} \\ \textcolor{red}{4.96\%},
    ] {6879 \\ 3.45\%};
    &
    \node (box22) [box_green,
    label=right:22641\\ \textcolor{forestgreen}{69.62\%} \\ \textcolor{red}{30.38\%},
    label=below:60592\\ \textcolor{forestgreen}{26.01\%} \\ \textcolor{red}{73.99\%},
    label=below right:199237\\ \textcolor{forestgreen}{74.05\%} \\ \textcolor{red}{25.95\%},
    ] {15762 \\ 7.91\%};
    \\
};
 \node [rotate=90,left=.05cm of conmat,anchor=center,text width=1.5cm] {\textbf{Predicted}};
\node [above=.05cm of conmat] {\textbf{Actual}};
\end{tikzpicture}}
		\caption{Data Augmentation Experiment (Section~\ref{data_augmentation_evaluation})} 
		\label{confusion_matrices_DA}
	\end{subfigure}
	\begin{subfigure}[b]{0.35\textwidth} 
		\scalebox{0.6}{\definecolor{forestgreen}{rgb}{0.0, 0.27, 0.13}

 \begin{tikzpicture}[align=center,
box/.style={draw,rectangle,minimum size=2cm,text width=1.5cm,align=left},
box_green/.style={draw,rectangle,fill=green!50,minimum size=2cm,text width=1.5cm,align=left},
box_red/.style={draw,rectangle,fill=red!40,minimum size=2cm,text width=1.5cm,align=left}]
\matrix (conmat) [row sep=.1cm,column sep=.1cm] {
    \node (box11) [box_green,
    label=left:\textbf{Non-Depressed },
    label=above:\textbf{\textbf{Non-Depressed }},
    ] {93176 \\ 66.37\%};
    &
    \node (box12) [box_red,
    label=above:\textbf{Depressed },
    label=above right:\textbf{Sum\_lin},
    label=right: 119909\\ \textcolor{forestgreen}{77.71\%} \\ \textcolor{red}{22.29\%},
    ] {26733 \\ 19.04\%};
    \\
    \node (box21) [box_red,
    label={left:\textbf{Depressed }},
    label=below left:\textbf{Sum\_col},
    label=below:99534\\ \textcolor{forestgreen}{93.16\%} \\ \textcolor{red}{6.39\%},
    ] {63658 \\ 4.53\%};
    &
    \node (box22) [box_green,
    label=right:22641\\ \textcolor{forestgreen}{69.62\%} \\ \textcolor{red}{30.38\%},
    label=below:40860\\ \textcolor{forestgreen}{34.57\%} \\ \textcolor{red}{65.43\%},
    label=below right:140394\\ \textcolor{forestgreen}{76.43\%} \\ \textcolor{red}{23.57\%},
    ] {14127 \\ 10.06\%};
    \\
};
 \node [rotate=90,left=.05cm of conmat,anchor=center,text width=1.5cm] {\textbf{Predicted}};
\node [above=.05cm of conmat] {\textbf{Actual}};
\end{tikzpicture}}
		\caption{Transfer Learning Experiment (Section~\ref{tranfer_learning_evaluation})} 
		\label{confusion_matrices_transfer}
	\end{subfigure}
	\begin{subfigure}[b]{0.35\textwidth} 
		\scalebox{0.6}{\definecolor{forestgreen}{rgb}{0.0, 0.27, 0.13}

 \begin{tikzpicture}[align=center,
box/.style={draw,rectangle,minimum size=2cm,text width=1.5cm,align=left},
box_green/.style={draw,rectangle,fill=green!50,minimum size=2cm,text width=1.5cm,align=left},
box_red/.style={draw,rectangle,fill=red!40,minimum size=2cm,text width=1.5cm,align=left}]
\matrix (conmat) [row sep=.1cm,column sep=.1cm] {
    \node (box11) [box_green,
    label=left:\textbf{Non-Depressed },
    label=above:\textbf{\textbf{Non-Depressed }},
    ] {167714 \\ 41.08\%};
    &
    \node (box12) [box_red,
    label=above:\textbf{Depressed },
    label=above right:\textbf{Sum\_lin},
    label=right: 301694\\ \textcolor{forestgreen}{55.59\%} \\ \textcolor{red}{44.41\%},
    ] {133980 \\ 32.82\%};
    \\
    \node (box21) [box_red,
    label={left:\textbf{Depressed }},
    label=below left:\textbf{Sum\_col},
    label=below:222452\\ \textcolor{forestgreen}{75.39\%} \\ \textcolor{red}{24.61\%},
    ] {54738 \\ 13.41\%};
    &
    \node (box22) [box_green,
    label=right:106521\\ \textcolor{forestgreen}{48.61\%} \\ \textcolor{red}{51.39\%},
    label=below:185763\\ \textcolor{forestgreen}{27.88\%} \\ \textcolor{red}{72.12\%},
    label=below right:408215\\ \textcolor{forestgreen}{53.77\%} \\ \textcolor{red}{46.23\%},
    ] {51783 \\ 12.69\%};
    \\
};
 \node [rotate=90,left=.05cm of conmat,anchor=center,text width=1.5cm] {\textbf{Predicted}};
\node [above=.05cm of conmat] {\textbf{Actual}};
\end{tikzpicture}}
		\caption{Generalization experiment (Section~\ref{generalization_evaluation})} 
		\label{confusion_matrices_generalization}
	\end{subfigure}
	\caption{Confusion Matrices of MFCC-based RNN generated on the test set of three different experiments: (a) Data Augmentation experiment, (b) Transfer learning experiment, (c) Generalization of the MFCC-based RNN model to other dataset experiment.} 
	\label{confusion_matrices}
\end{figure*}

\subsubsection{Transfer Learning evaluation}
\label{tranfer_learning_evaluation}

For depression assessment, Table.~\ref{table_transfer_binary} shows that the validation accuracy increased by 8.66 \% compared to the one obtained with the first experiment to reach 76.27\%. The validation drops by 0.1 and reaches 0.4.

As shown in Fig.~\ref{confusion_matrices_transfer} and Table.~\ref{table_transfer_confusion_matrix}, the F1 score of depression has increased by 8\% to reach 46\% compared to 38\% with data augmentation. This is explained by the 9\% increase of the recall of depression. 69\% of the samples predicted as depression are well-classified.\\ 

Pretraining the model over an independent, yet, related task has improved its ability to predict depression even without increasing the input data of depressed participants. By pretraining the MFCC-based RNN model on another dataset, it has learned more complex and abstract features in the first layers, which then improves its performance in the target task. 

\subsubsection{Gender effect assessment}
Throughout this experience, the input data is loaded into two separate groups: one for Male participants and one for Female participants. The purpose of this experience is to assess depression among each group. Thus, the MFCC-based RNN is trained for each one separately. The gender effect assessment is evaluated by comparing the Depressed/Non-Depressed accuracies for both genders. Initially, the validation accuracy for the Female group reaches 85\% with a validation loss of 0.32 compared to 83\% for Male with a validation loss of 0.34 (Table.~\ref{table_gender_effect}). The network is highly-performing in detecting depression for both genders. The validation accuracy increases by an average of 33.8\% compared to the initial results in the validation set with 67.61\% for depression assessment. Gender affects considerably the performances of the MFCC-based RNN in depression recognition and assessment. Adding another step in the proposed approach to recognize gender can improve the performances and it is to consider in future works. 

\subsubsection{Robustness to Noise}

The robustness to noise of the MFCC-based RNN framework is evaluated by adding noise to the baseline dataset used. Gaussian noise is added to the validation set which is 20\% of the input data. First, noise is added to a portion of 10\% of the validation data and then to the whole set. The gaussian noise is generated with a mean of 0 and a sigma of 0.1. As summarized in Table.~\ref{table_robustness_to_noise}, with 10\% of noise added to the validation set, the binary depression classification (depression/non-depression) accuracy drops by 8.27\% compared to the one obtained with transfer learning. The accuracy achieves 68\% which is 0.39\% higher than the baseline accuracy achieved with the MFCC-based RNN model before data augmentation and transfer learning. The drop of the overall accuracy is mainly caused by the 19\% drop of the depression F1 score that achieves 20\%. The F1 score of non-depression drops by 5\%  only to achieve 80\%. The performance of the deep framework proposed in this study is almost the same after adding 20\% of gaussian noise to the whole validation dataset. \\ 

When adding noise to the dataset, the MFCC-based RNN performances slightly decrease by 8\%. Its ability to well-classify depression decreases more than its ability to well-classify non-depression.  When increasing the noise within the validation data, the performance of the model stays stable and does not deteriorate more. The proposed network is robust to noise and performs even better than the baseline model trained without augmenting the data and without performing transfer of knowledge. 

\begin{table}[]
    \centering
    \begin{center}
        \begin{tabular}{|m{1.2cm}|m{2.2cm}|m{1.2cm}|m{0.75cm}|m{0.6cm}|}
            \hline
                &Class&Precision&Recall&F1\\
            \hline
                \multirow{3}{4em}{10\% of Gaussian noise} & Non-Depressed & 73\%  & 88\% & 80\% \\ 
                 & Depressed & 41\%  & 20\% & 27\% \\ 
                 & Accuracy &   &  & 68\% \\ 
            \hline
                \multirow{3}{4em}{20\% of Gaussian noise} &  Non-Depressed & 73\%  & 88\% & 80\% \\ 
                 & Depressed & 41\%  & 21\% & 27\% \\ 
                 & Accuracy &   &  & 68\% \\ 
            \hline
        \end{tabular}
    \end{center}
    \caption{MFCC-based RNN performances under depressed and non-depressed groups based on three metrics (precision, recall and FI score) after adding gaussian noise on a 20\% validation set.}
\label{table_robustness_to_noise}
\end{table}

\subsubsection{Generalization of the MFCC-based RNN model to other dataset}
\label{generalization_evaluation}
The proposed recurrent framework is tested on the AVi-D dataset. A good performance means that the MFCC-based RNN generalizes well to other datasets. The model is trained to classify the PHQ-8 binary while it is tested to classify the BDI-II binary, an other similar clinical test for depression assessment. The scores of the BDI-II test range differently from the scores of the PHQ-8 test. Therefore, the evaluation is performed only for binary classification of depression/non-depression. The model is tested over the AVi-D data of the two tasks Freeform and Northwind seperately. Next, the model is tested over the whole set of data of both tasks combined. \\

Table.~\ref{table_generalization} describes the final results of the generalization experiment. The best classification accuracy is achieved with the Freeform task which reaches 56\%. For both tasks combined, the confusion matrix in Fig.~\ref{confusion_matrices_generalization} shows that 12.69\% of the samples are correctly identified with depression. The depression F1 score drops by 11\% with the new dataset to reach 35\% for both tasks combined whereas it drops by 21\% with non-depression compared to the F1 scores obtained with transfer learning. \\

The classification accuracy of the generalization drops by 20.43\% while comparing it with the one achieved with transfer learning. However, the MFCC-based RNN model performs better in the generalization experiment in correctly identifying samples of depression: 12.69\% of the samples are correctly identified with depression while classifying the BDI-II binary. Whereas, only 10\% of the samples are correctly identified with depression with transfer learning. These results are explained by the different recording conditions of the participants speech. The Freeform and Northwind tasks are performed by actors who talk and sing. Depression is assessed differently with an other self-reported depression test with a different threshold level. 

\begin{table}[]
    \centering
    \begin{center}
        \begin{tabular}{|m{1.35cm}|m{2.31cm}|m{1.12cm}|m{0.75cm}|m{0.54cm}|}
            \hline
                Task&Class&Precision&Recall&F1\\
            \hline
                \multirow{3}{4em}{Freeform} & Non-Depression & 60\%  & 74\% & 66\% \\
                & Depression & 44\% & 30\% & 36\% \\
                & Accuracy & & & 56\% \\
            \hline
                \multirow{3}{4em}{Northwind} & Non-Depression & 49\%  & 78\% & 60\% \\ 
                & Depression & 55\% & 24\% & 34\% \\
                & Accuracy & & & 50\% \\
            \hline
                \multirow{3}{4em}{Both Tasks} & Non-Depression & 56\%  & 75\%  & 64\% \\ 
                & Depression & 49\% & 28\% & 35\% \\
                & Accuracy & & & 54\% \\
            \hline
        \end{tabular}
    \end{center}
    \caption{Generalization of the MFCC-based RNN model to other dataset. The MFCC-based RNN model is trained on DAIC-WOZ dataset and tested on AVi-D dataset. The task on the training concerns the PHQ-8 binary classification, while the task on the test concerns the BDI-II binary classification.}
\label{table_generalization}
\end{table}

 \subsubsection{Comparison with existing methods}
 
The MFCC-based Recurrent Neural Network is compared to a benchmark of works summarized in Table.\ref{baseline_results}. For predicting the depression severity levels, the proposed deep recurrent framework performs better than the architecture based on a Deep Convolutional Neural Network combined with a Deep Neural Network. The root mean square error (RMSE) achieved with the MFCC-based RNN for the PHQ-8 scores prediction is 0.168 while the best RMSE achieved with the DCNN-DNN is 1.46. The proposed framework is 9.75 times more performing than the DCNN-DNN. 
When it comes to predicting the PHQ-8 binary, the MFCC-based RNN performs 15\% better in detecting non-depression than the Convolutional Neural Network followed by a Long Short-Term Memory network. Yet, the latter is 4\% better in detecting depression.

\begin{table*}[t]
        \centering
        \begin{tabular}{|l|c|c|c|r|}
            \hline
                Method & Metrics & Audio Features & Model used & Performance\\
            \hline
                \multirow{2}{*}{\cite{yang2017hybrid}} & PHQ-8 scores & 238 LLD & DCNN + DNN & RMSE = 1.46 \\ 
                & & & & Depressed Male\\
            \hline
                \multirow{2}{*}{\cite{yang2017multimodal}} & PHQ-8 scores & 238 LLD & DCNN + DNN & RMSE = 5.59 \\
                & & & & Male \\
            \hline
                \multirow{2}{*}{\cite{ma2016depaudionet}} & PHQ-8 binary & Raw Audio Signal & CNN + LSTM & F1=70\% (Non Depression) \\
                & & + Mel Filter Bank & & F1=50\% (Depression)\\
            \hline
                \multirow{4}{*}{\textbf{MFCC-based RNN}} &  \multirow{4}{*}{\textbf{PHQ-8 scores}} & \multirow{4}{*}{\textbf{60 MFCC coefficients}} & \multirow{4}{*}{\textbf{LSTM}} & \multirow{4}{*}{\textbf{RMSE = 0.168}}\\
                & & & & \\
                & & & & \\
                & & & & \\
            \hline
                \multirow{4}{*}{\textbf{MFCC-based RNN} } &  \multirow{4}{*}{\textbf{PHQ-8 binary}} & \multirow{4}{*}{\textbf{60 MFCC coefficients}} & \multirow{4}{*}{\textbf{LSTM}} & \textbf{RMSE = 0.4}\\
                & & & & \textbf{Accuracy = 76.27\%}\\   
                & & & & \textbf{F1=85\% (Non Depression)}\\    
                & & & & \textbf{F1=46\% (Depression)}\\    
            \hline
        \end{tabular}
    \caption{Performance comparison of depression recognition and assessment methods on DAIC-WOZ dataset.}
\label{baseline_results}
\end{table*}

\subsubsection{Discussion} 
The MFCC-based RNN reaches an overall validation accuracy of 76.27\% in the depression assessment task with a root mean square error of 0.4. Two choices are made in building the proposed framework: the use of only the speech modality and the use of only the MFCC coefficients as features. This choice being guided by several criteria:
\begin{itemize}
    \item the high performance of MFCC coefficients in speech-related applications and frameworks,
    \item the robustness of the speech modality in automatic depression recognition,
    \item the non-invasiveness and the non-intrusion of the speech modality.
\end{itemize}
The proposed approach  presents good performance and small computational time. Considering other modalities like facial images can make the proposed framework intrusive but better performing. Adding other audio features could increase the computational time but it could improve the performances.
In this section, a study of the performances of a multi-modal and a multi-features frameworks is performed. The comparative study, as shown in Fig.~\ref{multi_modal_channel}, is elaborated to evaluate the impact of adding more features or other modalities on the model's performance.
\subparagraph{The multi-modal experiment:} In this experiment, as shown in Fig.~\ref{multi_modal}, visual features are aggregated with the deep audio features for automatic depression recognition. The added visual features used in the following experiments are provided by the DAIC-WOZ corpus. Twenty facial action units are indeed concatenated with the flattened output of the previously trained three LSTM layers. The concatenation vector of the newly introduced visual features and the extracted high-level MFCC features is then fed to the fully connected layers for classification. \\

The overall validation accuracy achieved marked an increase of 19.33\% to reach 95.6\%, while the validation loss decreased by 0.22 to reach 0.18. The depression F1 score doubled and reached 94\%, while the non-depression F1 score increased by 11\% to reach 96\%. \\ 
The F1 scores show that adding visual features highly improves the model's ability to recognize signs of depression. This concatenation has conquered the data imbalance issue leading to an increased ability to identify depression as much as it is able to identify non-depression.
\subparagraph{The multi-features or multi-channel experiment:} As shown in Fig.~\ref{multi_channel}, 53 extra acoustic features are used: the fundamental frequency (F0), the Voicing (VUV), the Normalized Amplitude Quotient (NAQ), the Quasi Open Quotient (QOQ), the Harmonic  difference (H1H2), the Parabolic Spectral Parameter (PSP), the Maxima Dispersion Quotient (MDQ), peakSlope, Rd, Rd\_conf,  the Harmonic Model and Phase Distortion Mean (HMPDM0-24), the Harmonic Model and Phase Distortion Deviations (HMPDD0-12), and the first five Formants. The flattened vector of acoustic features is concatenated with the flattened output of high-level MFCC features. The concatenated vector is then fed to the dense layers. \\

The overall accuracy reached 86\% on the validation set marking an increase of almost 10\%. Meanwhile, the validation RMSE reached 0.32 marking a decrease of 0.08. The non-depression F1 score increased by 5\% only, while the depression F1 score increased by 29\% to reach 75\%. These results go in line with state of the art. A shallow-based approach in \cite{williamson2016detecting} marked an increase of 12\% of the mean F1 score in a multi-channel experiment. Meanwhile, it marked an increase of 23\% in a multi-modal experiment. According to these results, it is interesting to add other features to better predict depression. Deep and handcrafted features are complementary. As more features are added to the high-level MFCC, the model gathers more information about signs of depression. It gains more identification knowledge. Adding more acoustic features increases performances by 10\% but the computational complexity increases. Extracting more features increases the prediction time cost. Adding visual features increases performances by 20\%, yet, it would be intrusive and invasive to the patients. Therefore, the application in real-time becomes more difficult and inconvenient.

\section{Conclusion and future works}
In this study, an MFCC-based Recurrent Neural Network is proposed to detect depression and to assess its severity levels. The audio recordings are preprocessed and the MFCC features are then extracted and normalized. The MFCC coefficients are fed to deep recurrent neural network of successive LSTM layers.
To overcome the lack of training data and overfitting problems, two approaches are considered: augmenting the training data and transferring knowledge from another related task. 
The proposed architecture is evaluated on the DAIC-WOZ corpus and promising results are achieved. 
For the future work, we plan to add a gender recognition step and to balance data classes in the proposed framework. A web-application could be also designed to automatically diagnose clinical depression without any medical assistance. 





\section*{References}
\bibliography{my_bib}

\begin{thebibliography}{10}
\expandafter\ifx\csname url\endcsname\relax
  \def\url#1{\texttt{#1}}\fi
\expandafter\ifx\csname urlprefix\endcsname\relax\def\urlprefix{URL }\fi
\expandafter\ifx\csname href\endcsname\relax
  \def\href#1#2{#2} \def\path#1{#1}\fi

\bibitem{greenberg2015economic}
P.~E. Greenberg, A.-A. Fournier, T.~Sisitsky, C.~T. Pike, R.~C. Kessler, The
  economic burden of adults with major depressive disorder in the united states
  (2005 and 2010), The Journal of clinical psychiatry 76~(2) (2015) 155--162.

\bibitem{mitchell2009clinical}
A.~J. Mitchell, A.~Vaze, S.~Rao, Clinical diagnosis of depression in primary
  care: a meta-analysis, The Lancet 374~(9690) (2009) 609--619.

\bibitem{dornaika2007inferring}
F.~Dornaika, B.~Raducanu, Inferring facial expressions from videos: Tool and
  application, Signal Processing: Image Communication 22~(9) (2007) 769--784.

\bibitem{wilkins2007inexpensive}
P.~Wilkins, T.~Adamek, N.~E. O’connor, A.~F. Smeaton, Inexpensive fusion
  methods for enhancing feature detection, Signal Processing: Image
  Communication 22~(7-8) (2007) 635--650.

\bibitem{scherer2015self}
S.~Scherer, G.~M. Lucas, J.~Gratch, A.~S. Rizzo, L.-P. Morency, Self-reported
  symptoms of depression and ptsd are associated with reduced vowel space in
  screening interviews, IEEE Transactions on Affective Computing~(1) (2015)
  59--73.

\bibitem{williamson2016detecting}
J.~R. Williamson, E.~Godoy, M.~Cha, A.~Schwarzentruber, P.~Khorrami, Y.~Gwon,
  H.-T. Kung, C.~Dagli, T.~F. Quatieri, Detecting depression using vocal,
  facial and semantic communication cues, in: Proceedings of the 6th
  International Workshop on Audio/Visual Emotion Challenge, ACM, 2016, pp.
  11--18.

\bibitem{yang2017hybrid}
L.~Yang, H.~Sahli, X.~Xia, E.~Pei, M.~C. Oveneke, D.~Jiang, Hybrid depression
  classification and estimation from audio video and text information, in:
  Proceedings of the 7th Annual Workshop on Audio/Visual Emotion Challenge,
  ACM, 2017, pp. 45--51.

\bibitem{valstar2016avec}
M.~Valstar, J.~Gratch, B.~Schuller, F.~Ringeval, D.~Lalanne, M.~Torres~Torres,
  S.~Scherer, G.~Stratou, R.~Cowie, M.~Pantic, Avec 2016: Depression, mood, and
  emotion recognition workshop and challenge, in: Proceedings of the 6th
  international workshop on audio/visual emotion challenge, ACM, 2016, pp.
  3--10.

\bibitem{lopez2014study}
P.~Lopez-Otero, L.~Dacia-Fernandez, C.~Garcia-Mateo, A study of acoustic
  features for depression detection, in: 2nd International Workshop on
  Biometrics and Forensics, IEEE, 2014, pp. 1--6.

\bibitem{cummins2011investigation}
N.~Cummins, J.~Epps, M.~Breakspear, R.~Goecke, An investigation of depressed
  speech detection: Features and normalization, in: Twelfth Annual Conference
  of the International Speech Communication Association, 2011.

\bibitem{tiwari2010mfcc}
V.~Tiwari, Mfcc and its applications in speaker recognition, International
  journal on emerging technologies 1~(1) (2010) 19--22.

\bibitem{ma2016depaudionet}
X.~Ma, H.~Yang, Q.~Chen, D.~Huang, Y.~Wang, Depaudionet: An efficient deep
  model for audio based depression classification, in: Proceedings of the 6th
  International Workshop on Audio/Visual Emotion Challenge, ACM, 2016, pp.
  35--42.

\bibitem{pampouchidou2017automatic}
A.~Pampouchidou, P.~Simos, K.~Marias, F.~Meriaudeau, F.~Yang, M.~Pediaditis,
  M.~Tsiknakis, Automatic assessment of depression based on visual cues: A
  systematic review, IEEE Transactions on Affective Computing (2017).

\bibitem{ringeval2018avec}
F.~Ringeval, B.~Schuller, M.~Valstar, R.~Cowie, H.~Kaya, M.~Schmitt,
  S.~Amiriparian, N.~Cummins, D.~Lalanne, A.~Michaud, et~al., Avec 2018
  workshop and challenge: Bipolar disorder and cross-cultural affect
  recognition, in: Proceedings of the 2018 on Audio/Visual Emotion Challenge
  and Workshop, ACM, 2018, pp. 3--13.

\bibitem{pampouchidou2017detection}
A.~Pampouchidou, O.~Simantiraki, C.-M. Vazakopoulou, K.~Marias, P.~Simos,
  Y.~Fan, F.~Meriaudeau, M.~Tsiknakis, D{\'e}tection de la d{\'e}pression par
  l’analyse de la g{\'e}om{\'e}trie faciale et de la parole, GRETSI, 2017.

\bibitem{jiang2017investigation}
H.~Jiang, B.~Hu, Z.~Liu, L.~Yan, T.~Wang, F.~Liu, H.~Kang, X.~Li, Investigation
  of different speech types and emotions for detecting depression using
  different classifiers, Speech Communication 90 (2017) 39--46.

\bibitem{low2010detection}
L.-S.~A. Low, N.~C. Maddage, M.~Lech, L.~B. Sheeber, N.~B. Allen, Detection of
  clinical depression in adolescents’ speech during family interactions, IEEE
  Transactions on Biomedical Engineering 58~(3) (2010) 574--586.

\bibitem{jiang2018detecting}
H.~Jiang, B.~Hu, Z.~Liu, G.~Wang, L.~Zhang, X.~Li, H.~Kang, Detecting
  depression using an ensemble logistic regression model based on multiple
  speech features, Computational and mathematical methods in medicine 2018
  (2018).

\bibitem{ringeval2015avec}
F.~Ringeval, B.~Schuller, M.~Valstar, R.~Cowie, M.~Pantic, Avec 2015: The 5th
  international audio/visual emotion challenge and workshop, in: Proceedings of
  the 23rd ACM international conference on Multimedia, ACM, 2015, pp.
  1335--1336.

\bibitem{alghowinem2013comparative}
S.~Alghowinem, R.~Goecke, M.~Wagner, J.~Epps, T.~Gedeon, M.~Breakspear,
  G.~Parker, A comparative study of different classifiers for detecting
  depression from spontaneous speech, in: 2013 IEEE International Conference on
  Acoustics, Speech and Signal Processing, IEEE, 2013, pp. 8022--8026.

\bibitem{ringeval2017avec}
F.~Ringeval, B.~Schuller, M.~Valstar, J.~Gratch, R.~Cowie, S.~Scherer,
  S.~Mozgai, N.~Cummins, M.~Schmitt, M.~Pantic, Avec 2017: Real-life
  depression, and affect recognition workshop and challenge, in: Proceedings of
  the 7th Annual Workshop on Audio/Visual Emotion Challenge, ACM, 2017, pp.
  3--9.

\bibitem{valstar2013avec}
M.~Valstar, B.~Schuller, K.~Smith, F.~Eyben, B.~Jiang, S.~Bilakhia,
  S.~Schnieder, R.~Cowie, M.~Pantic, Avec 2013: the continuous audio/visual
  emotion and depression recognition challenge, in: Proceedings of the 3rd ACM
  international workshop on Audio/visual emotion challenge, ACM, 2013, pp.
  3--10.

\bibitem{trigeorgis2016adieu}
G.~Trigeorgis, F.~Ringeval, R.~Brueckner, E.~Marchi, M.~A. Nicolaou,
  B.~Schuller, S.~Zafeiriou, Adieu features? end-to-end speech emotion
  recognition using a deep convolutional recurrent network, in: 2016 IEEE
  international conference on acoustics, speech and signal processing (ICASSP),
  IEEE, 2016, pp. 5200--5204.

\bibitem{yang2017multimodal}
L.~Yang, D.~Jiang, X.~Xia, E.~Pei, M.~C. Oveneke, H.~Sahli, Multimodal
  measurement of depression using deep learning models, in: Proceedings of the
  7th Annual Workshop on Audio/Visual Emotion Challenge, ACM, 2017, pp. 53--59.

\bibitem{janse2014comparative}
P.~V. Janse, S.~Magre, P.~Kurzekar, R.~Deshmukh, A comparative study between
  mfcc and dwt feature extraction technique, International Journal of
  Engineering Research and Technology 3~(1) (2014) 3124--3127.

\bibitem{gratch2014distress}
J.~Gratch, R.~Artstein, G.~M. Lucas, G.~Stratou, S.~Scherer, A.~Nazarian,
  R.~Wood, J.~Boberg, D.~DeVault, S.~Marsella, et~al., The distress analysis
  interview corpus of human and computer interviews., in: LREC, Citeseer, 2014,
  pp. 3123--3128.

\bibitem{livingstone2018ryerson}
S.~R. Livingstone, F.~A. Russo, The ryerson audio-visual database of emotional
  speech and song (ravdess): A dynamic, multimodal set of facial and vocal
  expressions in north american english, PloS one 13~(5) (2018) e0196391.

\bibitem{ouyang2017audio}
X.~Ouyang, S.~Kawaai, E.~G.~H. Goh, S.~Shen, W.~Ding, H.~Ming, D.-Y. Huang,
  Audio-visual emotion recognition using deep transfer learning and multiple
  temporal models, in: Proceedings of the 19th ACM International Conference on
  Multimodal Interaction, ACM, 2017, pp. 577--582.

\bibitem{atalay2018comparison}
T.~Atalay, D.~Ayata, Y.~Yaslan, Comparison of feature selection methods in
  voice based emotion recognition systems, in: 2018 26th Signal Processing and
  Communications Applications Conference (SIU), IEEE, 2018, pp. 1--4.

\bibitem{pham2018end}
H.~X. Pham, Y.~Wang, V.~Pavlovic, End-to-end learning for 3d facial animation
  from speech, in: Proceedings of the 2018 on International Conference on
  Multimodal Interaction, ACM, 2018, pp. 361--365.

\bibitem{valstar2014avec}
M.~Valstar, B.~W. Schuller, J.~Krajewski, R.~Cowie, M.~Pantic, Avec 2014: the
  4th international audio/visual emotion challenge and workshop, in:
  Proceedings of the 22nd ACM international conference on Multimedia, ACM,
  2014, pp. 1243--1244.

\end{thebibliography}
\bibliographystyle{elsarticle-num}


\end{document}